\documentclass[acmtog,authorversion]{acmart}
\acmSubmissionID{241}

\usepackage{booktabs} 
\usepackage{wrapfig}
\usepackage{svg}
\usepackage{subfig}
\usepackage[ruled]{algorithm2e} 

\SetAlFnt{\small}
\SetAlCapFnt{\small}
\SetAlCapNameFnt{\small}
\SetAlCapHSkip{0pt}

\usepackage{xspace}
\makeatletter
\DeclareRobustCommand\onedot{\futurelet\@let@token\@onedot}
\def\@onedot{\ifx\@let@token.\else.\null\fi\xspace}
\def\eg{\emph{e.g}\onedot} 
\def\ie{\emph{i.e}\onedot}

\def\wrt{w.r.t\onedot} 
\def\etal{\emph{et al}\onedot}
\makeatother

\newcommand{\f}{{\mathbf f}}

\newcommand{\bM}{\mathbf{M}}

\newcommand{\bff}{\mathbf{f}}

\newcommand{\bn}{\mathbf{n}}

\newcommand{\bv}{\mathbf{v}}
\newcommand{\bx}{\mathbf{x}}

\newcommand{\bZero}{\mathbf{0}}

\newcommand{\tran}{\mathsf{T}}
\newcommand{\dd}{\mathrm{d}}

\newcommand{\new}[1]
{
\textcolor{black}{#1}
}

\begin{document}
\title{No Free Slide: Spurious Contact Forces in Incremental Potential Contact}

\author{Yinwei Du}
\affiliation{%
  \institution{ETH Z{\"u}rich}
  \country{Switzerland}
}
\email{yinwei.du@inf.ethz.ch}

\author{Yue Li}
\affiliation{%
  \institution{ETH Z{\"u}rich}
  \country{Switzerland}
}
\email{yue.li@inf.ethz.ch}

\author{Stelian Coros}
\affiliation{%
  \institution{ETH Z{\"u}rich}
  \country{Switzerland}
}
\email{stelian@inf.ethz.ch}

\author{Bernhard Thomaszewski}
\affiliation{%
  \institution{ETH Z{\"u}rich}
  \country{Switzerland}
}
\email{bthomasz@ethz.ch}


\begin{abstract}
Modeling contact between deformable solids is a fundamental problem in computer animation, mechanical design, and robotics. 
Existing methods based on $C^0$-discretizations---piece-wise linear or polynomial surfaces---suffer from discontinuities and irregularities in tangential contact forces, which can significantly affect simulation outcomes and even prevent convergence. To overcome this limitation, we employ smooth surface representations for both contacting bodies. Through a series of test cases, we show that our approach offers advantages over existing methods in terms of accuracy and robustness for both forward and inverse problems. The contributions of our work include identifying the limitations of existing methods, examining the advantages of smooth surface representation, and proposing forward and inverse problems to analyze contact force irregularities.
\end{abstract}
%




\maketitle
\section{Introduction}
Modeling contact between deformable objects is a fundamental problem in science and engineering. It is at the heart of many applications in computer animation, mechanical design, and robotics.
The computer graphics community has made great strides in contact detection and response over the past two decades. 
The state-of-the-art is perhaps best reflected in the Incremental Potential Contact (IPC) method by Li \etal \shortcite{li2020incremental} that combines fail-safe collision detection with implicit integration of contact forces based on log-barrier penalty functions. This algorithm is able to generate compelling and stable animations for highly challenging contact scenarios.
Nevertheless, there are still limitations with existing formulations that demand further investigation. Here we draw attention to a particular and fundamental problem, \ie, the discontinuities and irregularities in tangential contact forces induced by non-smooth surface discretization.
\begin{figure}
    \centering
    \includegraphics[width=\linewidth]{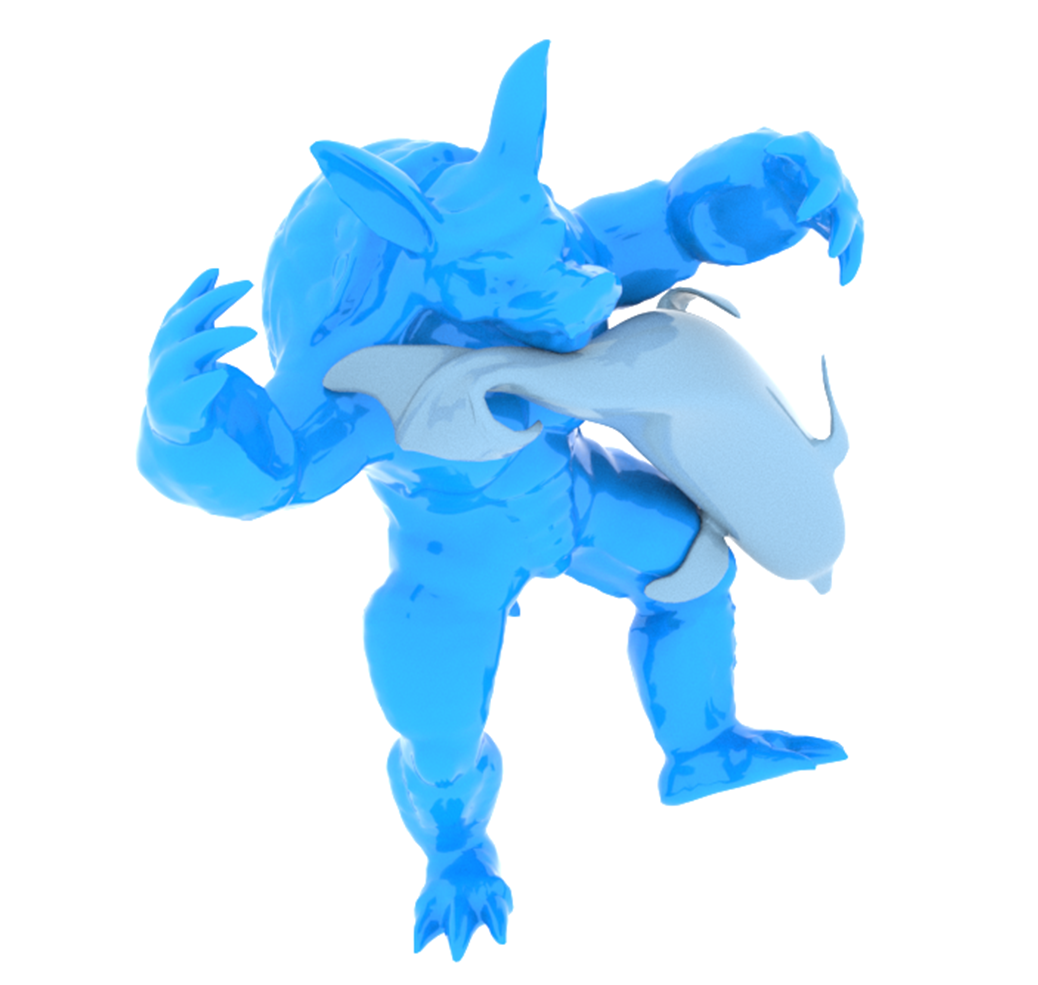}
    \caption{Catch-of-the-day. \new{We use smooth surface representations to accurately resolve contact between deformable objects such as these soft armadillo and dolphin models. 
    } }
    \label{fig:my_label}
\end{figure}
In engineering, arguably the most widely used approach for handling contact between deformable bodies is the Node-to-Segment (NTS) method \cite{ZAVARISE20093428}. For each vertex (\textit{node}) of a given discretized surface, the NTS method first finds the closest triangle (\textit{segment}) on the other surface. Contact forces are then determined based on these vertex-triangle pairs. Since collision-avoiding forces have to increase or maintain the minimum distance between collision pairs, they must align with the triangle's normal. Consequently, when a contact point migrates from one triangle to another, contact force directions change abruptly, and these discontinuities are highly problematic for implicit solvers.
IPC addresses this problem using smoothly-clamped penalty functions that allow multiple collision pairs for the same nodes to be active simultaneously. While there are no contact force discontinuities with this approach, our analysis shows that the superposition of per-primitive potentials leads to energy walls at element transitions that resist tangential motion, even for the perfectly planar, frictionless case. These erroneous forces degrade the accuracy of forward simulations and can even prevent convergence for inverse problems.

\new{In this work, we show that an alternative approach for} deformable contact modeling based on smooth surface representations eliminates contact force discontinuities and irregularities. Our method builds on Implicit Moving Least Squares (IMLS), a mesh-less representation that describes smooth surfaces as the zero levelset of an implicit function defined through position and normal data.
As we show through a series of test cases, \new{this} smooth representation offers advantages over existing methods based on $C^0$-discretizations for both forward and inverse problems. 

In summary, our work makes the following contributions:
\begin{itemize}
\item We identify a fundamental limitation of contact handling methods for deformable bodies that rely on $C^0$-discretizations. In particular, we show that spurious tangential forces can significantly affect the relative motion between contacting surfaces. \new{We furthermore demonstrate that contact handling with smooth surface representations eliminates force discontinuities and other irregularities.
}
\item We propose a series of forward and inverse problems to analyze and quantify contact force irregularities. Using these tests, we show that existing methods based on $C^0$-discre\-ti\-za\-tions can significantly alter simulation outcomes and may even prevent convergence while a smooth formulation eliminates these problems.
\end{itemize}

\section{Related Work}

\paragraph{Contact Modeling in Graphics}
Handling contact is a fundamental problem in graphics. 
Early research primarily focused on contacts between rigid bodies~\cite{baraff1991coping,moore1988collision,mirtich1995impulse,kaufman2005fast,kaufman2008staggered,kry2003continuous}. 
To handle contacts between deformable bodies, most approaches relied on impulses, i.e., velocity-level corrections for contacting primitive pairs~\cite{bridson2002robust,harmon2008robust}. 
However, as a post process to time stepping, these impulses do not result in equilibrium states.
%
Instead of using quadratical penalty potentials, Harmon \etal~\shortcite{Harmon2009} proposed an asynchronous time stepping strategy with layered collision potentials, which are infinitely stiff in the limit. While accurate and robust, this explicit time stepping requires extremely long computation times. 
More recently, Li et al.~\shortcite{li2020incremental} introduce a fully implicit treatment of contact using smoothly clamped barrier functions. This incremental potential contact (IPC) paradigm has been generalized to handle co-dimensional objects~\cite{li2020codimensional}, rigid bodies~\cite{ferguson2021intersection,lan2022affine}, and reduced models~\cite{lan2021medial}. 
While IPC enjoys significantly improved robustness, we show that the superposition of per-element potentials leads to spurious tangential forces that can significantly affect simulation outcomes.

\paragraph{Frictional Contact}
Accurate modeling of friction is of paramount importance in many applications and has been extensively studied in the graphics community for both forward modeling~\cite{daviet2020simple,verschoor2019efficient,li2018implicit,lfp21} and inverse design~\cite{li2022diffcloth,geilinger2020add}.
However, the inherently non-smooth nature of the governing Maximal Dissipation Principle~\cite{goyal1991planar,moreau2011unilateral} poses significant challenges for conventional Newton-type integrators. 
Consequently, a variety of customized second-order solvers have been developed, such as the Non-smooth Newton solvers ~\cite{alart1991mixed,Daviet2011,macklin2019non} and different iterative strategies~\cite{jean1992unilaterality,kaufman2008staggered,Otaduy2009,kaufman2014adaptive}.
While our work does not consider friction explicitly,  accurately modeling normal forces is a prerequisite for meaningful frictional behavior. We show that even in frictionless simulations, spurious tangential forces can lead to a non-negligible macroscopic friction effect.
\paragraph{Contacts with Implicit Representation}
Implicit representations have found widespread applications in various areas of computer graphics~\cite{turk2002modelling,shen2004interpolating} and have emerged as a valuable tool for handling contact.
Previous work has leveraged signed distance fields (SDFs)~\cite{jones20063d,frisken2000adaptively,koschier2017hp} for contact handling between deformable objects~\cite{fisher2001deformed,gascuel1993implicit} and skinning techniques~\cite{mcadams2011efficient,vaillant2013implicit,vaillant2014robust}.
Macklin \etal~\shortcite{macklin2020local} further improve the robustness of this approach by replacing point-based sampling with a method that finds the closest point between the SDF isosurface and a given mesh vertex.
However, one limitation of SDFs defined on polygonal meshes is their potential discontinuity across element boundaries, unless parallel edges are specifically considered~\cite{li2020codimensional}.
Inspired by Larionov \etal~\shortcite{lfp21}, we use Implicit Moving Least Squares (IMLS) ~\cite{levin2004mesh,kolluri2008provably,oztireli2009feature} to construct $C^2$ continuous surfaces whose smooth normal fields eliminate contact force irregularities.
%
We show through experiments that this approach leads to higher accuracy for both forward simulation and inverse design. 
%
%

\paragraph{Contact Problems in Engineering}
Contact problems have been extensively studied by the engineering community~\cite{Wriggers04Computational,popov2019handbook}.
%
Perhaps the most widely used method is the Node-to-segment (NTS) approach~\cite{hughes1976finite,ZAVARISE20093428}.  
However, the contact force discontinuities arising from simple Node-to-Segment methods are well-known and alternatives have been explored. 
Using an integral formulation that extends over the entire region of contact, Mortar methods~\cite{maday1988nonconforming,bernardi1989new} are arguably the most accurate approaches that exist today. We refer to the work by De Lorenzis \etal~\cite{DeLorenzis17Computational} for a comparison between the NTS, Mortar methods, and other approaches.
For all its accuracy and robustness, the complexity of Mortar methods is already daunting for forward simulation due to the requirement for meshing the gap between contact regions. Integrating this approach into an inverse problem solver seems all but infeasible. 
Despite the different choices in contact handling strategy, finite elements  remain the standard discretization for deformable solids. 
While we use finite elements for modeling soft body dynamics, we augment the underlying mesh representation with a smooth surface for contact resolution.
We show that the smooth distance fields offered by meshless representations greatly simplifies computation and implicit integration of contact forces.

\section{Deformable Contact Mechanics}
We focus on solving the dynamics of deformable solids with contact in a frictionless setting. We cast the governing equations into an unconstrained optimization problem and model contacts using penalty functions. Finding dynamic equilibrium states then amounts to minimizing the potential
\begin{equation} \label{eq::dynamics}
   E(\bx) = \frac{1}{2} \dot\bx^{\tran} \bM \dot\bx + \Psi(\bx) + \bx^{\tran} \f_e + E_{\mathrm{contact}}(\bx)\ ,
\end{equation}
where $\bx$ denote nodal positions, $\bM$ is the mass matrix, $\Psi(\bx)$ is the elastic potential and $\bff_e$ are external force.
Using fully implicit Euler for time integration, we obtain end-of-step positions $\bx_{t+1}$ by solving the problem
\begin{align}\label{eq::timestep}
\begin{split}
     \bx_{t+1} = \arg \min_{\bx} E(\bx) &= \frac{1}{2h^2} \left(\bx^{\tran} \bM \bx - \bx^{\tran}\bM(\bx_t + \bv_th)\right) \\
     &+ \Psi(\bx) + \bx^{\tran} \f_e + E_\mathrm{contact}(\bx),
\end{split}
\end{align}
where $h$ is the step size, and $\bx_t$ and $\bv_t$ are current positions and velocities, respectively.

\paragraph{Contact Constraint} 
The geometrical constraint between two deformable surfaces $\mathcal{B}_1$ and $\mathcal{B}_2$ is described through a measure for proximity known as the \textit{gap function} $g_n(\bx)$, which is defined as
\begin{equation} \label{eq::eq4}
    g_n(\bx^{(1)}) = -\bn_c\cdot(\bx^{(1)}-\hat{\bx}^{(2)}) \ ,
\end{equation}
where $\bx^{(1)}$ denote points on $\mathcal{B}_1$, $\hat{\bx}^{(2)}$ are corresponding contact points on $\mathcal{B}_2$, and $\bn_c$ are normal vectors. A standard choice for computing $\hat{\bx}^{(2)}$ is to use closest point projection,
\begin{equation} \label{eq::eq5}
    \hat{\bx}^{(2)} = \underset{\bx^{(2)} \in \mathcal{B}_2}{\arg\min} \ ||\bx^{(1)}-\bx^{(2)}|| \ .
\end{equation}
%
We impose the non-intersection constraint $g_n(\bx^{(1)})\geq 0$ with unilateral penalty functions $E_{\mathrm{contact}}$. 
It is straightforward to design $C^2$-continuous potentials as long as the gap function is continuous. However, preserving this smooth property becomes challenging when working with piece-wise linear discretizations, which exhibit discontinuities at element boundaries. While IPC constructs a smooth contact potential even for piece-wise linear meshes, as demonstrated in the subsequent section, there remains a fundamental limitation with this approach.
 
\section{Contact Energies on Piece-wise Linear Surfaces}
Piece-wise linear surfaces are arguably the most commonly used representation for deformable objects.
However, constructing contact forces with sufficient smoothness and regularity proves particularly challenging in this discrete setting.
In this section, we first demonstrate why early approaches, such as the Node-to-Segment method, fail to satisfy the smoothness requirements of gradient-based optimization methods (Sec.~\ref{sec:NTS}). Then we provide a detailed discussion on the strengths and limitations of IPC (Sec.~\ref{sec:IPC}).
\subsection{Node-to-Segment Method}
\label{sec:NTS}
\begin{figure}[h]
    \centering
    \includegraphics[width = 0.6\linewidth]{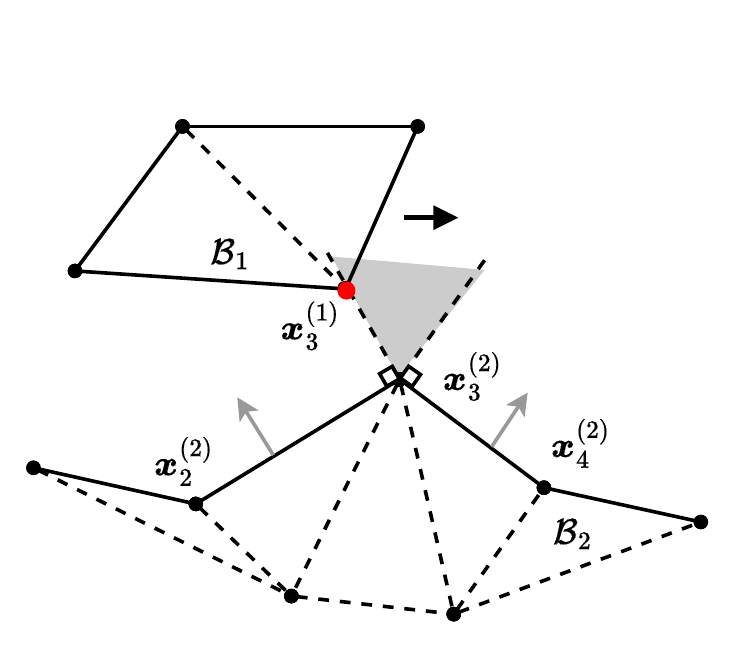}
    \caption{Discontinuity across element boundaries. As the red vertex moves in the indicated direction, stepping into or out of the gray region leads to discontinuities in contact forces.}
    \label{fig:IssueWithNTSProjection}
\end{figure}

The Node-to-Segment (NTS) method operates on a straightforward principle: it first identifies the closest segment for a given node and then uses the distance between the node and segment to define the gap function. Whenever the gap function yields a sufficiently small value, a penalty function is instantiated such that the corresponding contact force, collinear with the segment's normal, prevents intersection.
Fig.~\ref{fig:IssueWithNTSProjection} illustrates a scenario where two surfaces $\mathcal{B}_1$ and $\mathcal{B}_2$ are in close contact, with surface $\mathcal{B}_1$ moving in a given direction as indicated.
To compute the gap function for contact resolution, it is necessary to identify pairs of contact primitives and calculate the distance between them. Consider a vertex on surface $\mathcal{B}_1$ (shown in red). As the vertex enters or exits the shaded region where it is equidistant to both edges, the closest edge on surface $\mathcal{B}_2$ changes. 
Consequently, the contact force experiences an abrupt change in direction.
This force discontinuity occurs whenever a given node is at the same closest distance to multiple segments. This discrete change in contact primitive selection thus presents a fundamental problem of the NTS method.
\subsection{Incremental Potential Contact}
\label{sec:IPC}
The incremental potential contact method uses smoothly-clamped barrier functions to enforces non-intersection constraints between close primitive pairs, \eg, vertex-triangle, and edge-edge pairs. This involves finding a set of primitives in close proximity. 
Whereas the NTS method selects a single segment for each vertex, IPC allows multiple primitive pairs for the same vertex to be active simultaneously. The collision response for any given vertex is then given as the superposition of per-primitive potentials. Removing the need to make discrete decisions eliminates an important source of non-smoothness. As explained next, however, the superposition principle also induces energy variations when multiple per-primitives become active. 
%

\subsection{Spurious Tangential Forces in IPC}

Both IPC and NTS methods approximate the gap function between two contacting surfaces using per-primitive distances. However, IPC avoids ambiguities in closest-point projection by considering multiple active collision primitives for a given vertex. For instance, the contact energy for point $\bx^{(1)}_{3}$ in Fig.~\ref{fig:IssueWithNTSProjection} is computed as 
\begin{align}
\begin{split}
     E_{IPC} &= \kappa \sum_{i \in C} E_i \\ &=\kappa ( \phi(\bx^{(1)}_{3}, \bx^{(2)}_{2}\bx^{(2)}_{3}) + \phi(\bx^{(1)}_{3}, \bx^{(2)}_{3}\bx^{(2)}_{4}) \\ &+ \phi(\bx^{(1)}_{3}, \bx^{(2)}_{3})) \ ,
\end{split}
\end{align}
where $\phi(\bx_i, \bx_j\bx_k)$ represents the barrier energy between point $\bx_i$ and line segment $\bx_j\bx_k$ and $\phi(\bx_i, \bx_j)$ represents the energy between two points $\bx_i$ and $\bx_j$.
\begin{figure}[h]
    \centering
    \includegraphics[width=\linewidth]{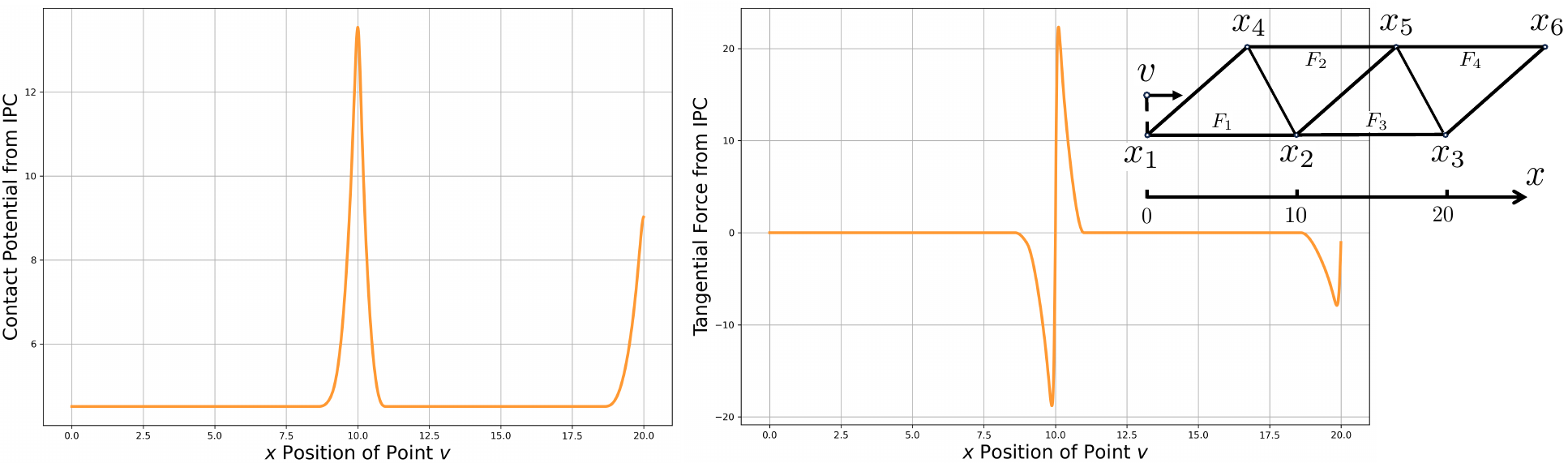}
    \caption{Energy walls. We plot the contact potential (left) and the tangential force (right) from IPC as a single vertex slides on a plane discretized with triangle elements (\textit{top right}). IPC allows for multiple per-primitive potentials to be active concurrently (\eg at $x = x_2$ or $x = x_3$), leading to energy walls, \ie, local maxima in contact energy whose gradients resist tangential motion.}
    \label{fig:IPCEnergyWall}
\end{figure}
\par
With this superposition approach, IPC yields smooth contact potentials and thus eliminates force discontinuities.
Unfortunately, the superposition of per-primitive potentials leads to undesirable \textit{energy walls} when a vertex is close to several primitives. We illustrate this problem using a simple setting (Fig.\ref{fig:IPCEnergyWall}), where we examine the contact potential from IPC between a vertex $v$ and a triangulated plane. We assume that the initial distance between the vertex and plane is smaller than the support radius of the barrier function, resulting in a nonzero contact potential.
As $v$ moves parallel to the plane, the constraint set---the set of \textit{active} primitive pairs---evolves from a single element (face $F_1$) to three elements (faces $F_1$, $F_2$, and $F_3$), and to two elements (faces $F_3$ and $F_4$).
As can be seen from Fig.~\ref{fig:IPCEnergyWall}, since IPC superimposes  contact potentials from all elements in the constraint set, the resulting energy is not constant. Rather, we observe energy peaks at $x= x_2 =10$ and $x = x_3 = 20$ where contributions from multiple elements amplify to a maximum degree. 
These \textit{energy walls} give rise to spurious tangential forces since the vertex has to perform mechanical work to keep moving at a constant distance. These tangential forces affect simulation outcomes and may impede progress for inverse problems as we show in Sec. \ref{sec:result}.
\section{Deformable Contact with Smooth Representations}
The problems encountered with NTS and IPC approaches can be attributed to the inherent limitations of piece-wise linear (or polynomial) surface representations. It is worth noting that using elements with higher polynomial order alone does not resolve this problem unless inter-element continuity is explicitly enforced. 
Rather, addressing these problems necessitates a fully $C^1$-continuous surface representation. While mesh-based representations based on, e.g., subdivision surfaces are a viable option \cite{Montes20Skintight}, we focus on meshless representations based on Implicit Moving-Least Squares (IMLS). 
%
IMLS enables the construction of a smooth signed distance field from a given input mesh. In the following, we first provide a brief summary of IMLS (Sec.~\ref{sec:basics}) and then show how to leverage this smooth representation to resolve contact between deformable bodies (Sec.~\ref{sec:IMLS_contact}).
\subsection{IMLS Basics}
\label{sec:basics}
\begin{figure}\centering
\includegraphics[width=0.9\linewidth]{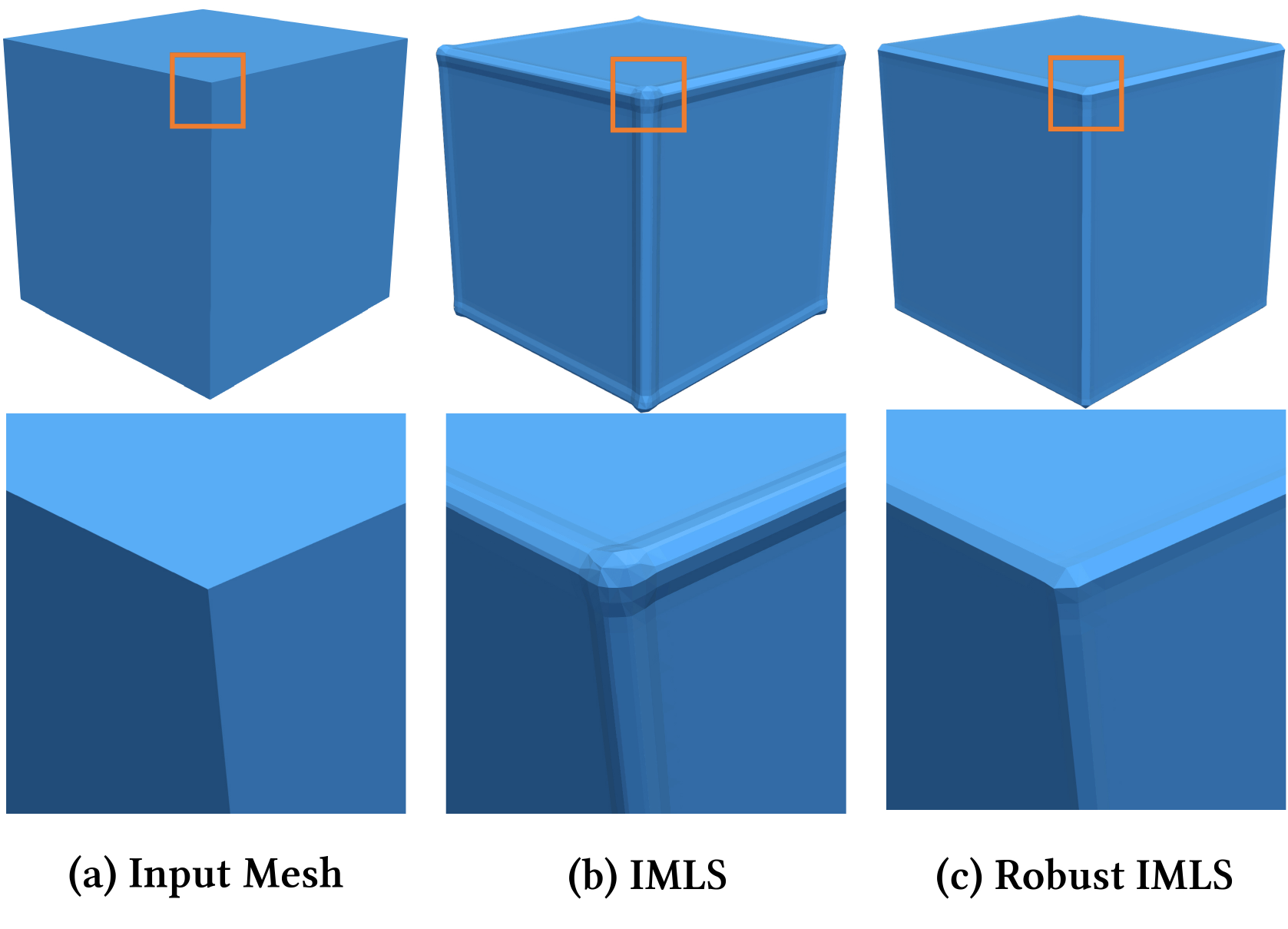}
\caption{Surface reconstruction. As can be seen from the close-up view in the second row, the standard IMLS formulation leads to over-smoothed reconstructions (b) of the input geometry (a). The robust IMLS method demonstrates better preservation of sharp features (c).}
\label{fig:RIMLSandIMLS}
\end{figure}

Given a set of points $\bx_i \in \mathbb{R}^d$ and their normals $\bn_i$, the corresponding IMLS surface is defined implicitly as the zero level set 
\begin{equation} \label{eq::eq12}
  f(\bx) = \frac{\sum \bn^{\tran}_i (\bx-\bx_i)\phi_i(\bx)}{\sum \phi_i(\bx)} = 0,
\end{equation}
where $\phi_i(\bx)$ are locally-supported, smoothly clamped radial basis functions that can be evaluated at any spatial location $\bx$ \cite{kolluri2008provably}.
One limitation of conventional IMLS formulation is its inability to preserve geometric features. This can be observed in Figure~\ref{fig:RIMLSandIMLS}(b), where a cube mesh is reconstructed using Eqn. \ref{eq::eq12} leading to over-smoothing around edges and corners.
Oztireli \etal~\shortcite{oztireli2009feature} addressed this limitation using robust local kernel regression. Iteratively re-weighted least squares minimization is used to obtain the implicit surface $f^k(\bx)$
\begin{equation} \label{eq::eq13}
  f^k(\bx) = \underset{s_0}{\arg\min} \sum (s_0+(\bx_i-\bx^{\tran})\bn_i)^2\phi_i(\bx) w(r^{k-1}_i)\ ,
\end{equation}
where $r^{k-1}_i$ is the residual between the $(k-1)$-th iteration $f^{k-1}(\bx) $ and $(\bx_i-\bx)^{\tran} \bn_i$. Oztireli \etal further point out that samples belonging to different sides of sharp features should not be interpreted as outliers, while their normals should. This observation inspires the use of a new weighting term $w_n$ that penalizes samples whose normals significantly differ from the reconstructed IMLS surface. 
The final robust IMLS surface is defined as
\begin{equation} \label{eq::eq14}
  f^k(\bx) = \frac{\sum \bn_i^{\tran}(\bx_i-\bx^{\tran})\phi_i(\bx) w(r^{k-1}_i) w_n(\Delta\bn_i^{k-1})}{\sum \phi_i(\bx) w(r^{k-1}_i) w_n(\Delta\bn_i^{k-1})} ,
\end{equation}
where $\Delta\bn_i^{k-1} = ||\nabla f^k(\bx)-\bn_i||$ measures the difference between the normal at sample point $\bn_i$ and the gradient of the IMLS surface at query point $\bx$.
As shown in Fig. \ref{fig:RIMLSandIMLS}(c), the robust IMLS formulation leads to improved preservation of sharp features.
%
Using this IMLS formulation, the corresponding signed distance function $f(\bx)$  is continuously differentiable when the basis functions $\phi(\bx)$ are continuously differentiable~\cite{levin1998approximation}. For notational convenience, we use $\psi(\bx)$ instead of $f^k(\bx)$ to represent the signed distance function in the remaining discussion.

\subsection{Smooth Contact Potential}
\label{sec:IMLS_contact}
With the IMLS formulation in place, we can now proceed to the definition of smooth contact potentials. 
We use solid finite elements for modeling the mechanics of deformable objects. We construct IMLS surfaces directly from the corresponding surface meshes using their vertex positions and normals. 
Our method largely follows the approach by Larionov \etal~\shortcite{lfp21}, but we introduce a few important modifications. Rather than using a cubic weighting function for simplification, we directly use the surface level set function (Eqn.~\ref{eq::eq14}) to evaluate the distance. Furthermore, instead of using face normals, we employ vertex normals computed from area-weighted face normals. Finally, instead of using hard constraints and off-the-shelf nonlinear programming solvers, we use unilateral penalty functions allowing us to leverage efficient second-order solvers for unconstrained minimization problems. 
Given a set of deformable objects $\mathcal{B}$ with positions and normals denoted as $\bx$ and $\bn$, respectively, we construct the contact potential
\begin{equation}
    E_\mathrm{contact} = \sum_{i \in \mathcal{B}} \sum_{j \in \mathcal{B}, j \neq i} b\left(\psi_j(\bx_i, \bn_i(\bx_i))\right),
\end{equation}

where $\psi_j$ is the IMLS surface of object $j$ and 
$b$ is a smooth \begin{wrapfigure}{r}{0.25\textwidth} 
    \centering
    \includegraphics[width=\linewidth]{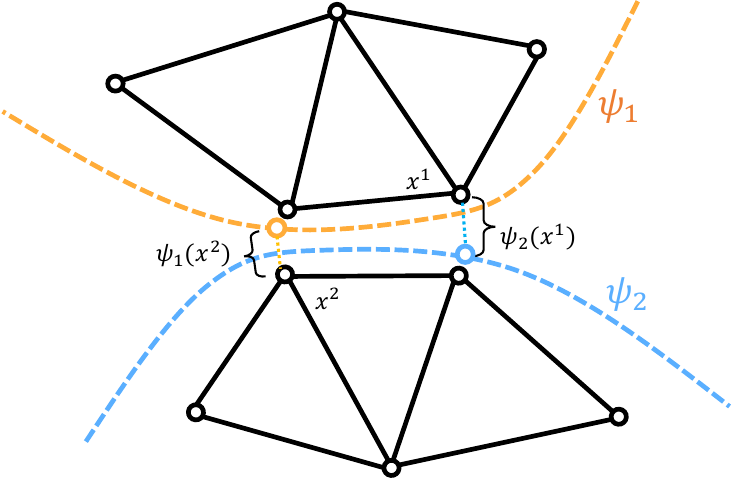}
    \label{fig:my_label}
\end{wrapfigure}
unilateral penalty function that becomes active when $\psi_j$ is smaller than a threshold value. The schematic figure illustrating a 2D example can be seen in the inset.
To simulate contact between deformable bodies represented by IMLS surfaces, we simply replace the contact potential in Eqn.~\ref{eq::dynamics} with the above expression. 

\section{Results}
We compare the behavior of IPC and IMLS-based formulations on a set of experiments that include both qualitative and quantitative benchmarks for forward and inverse problems (Sec.~\ref{sec:compare_IPC}). We furthermore demonstrate that smooth representations based on robust IMLS generalize to complex geometries with intricate collisions (Sec.~\ref{sec:result3d}).
\label{sec:result}

\subsection{Comparison with IPC} 
\label{sec:compare_IPC}
\begin{figure}[h]
    \centering
    \includegraphics[width = \linewidth]{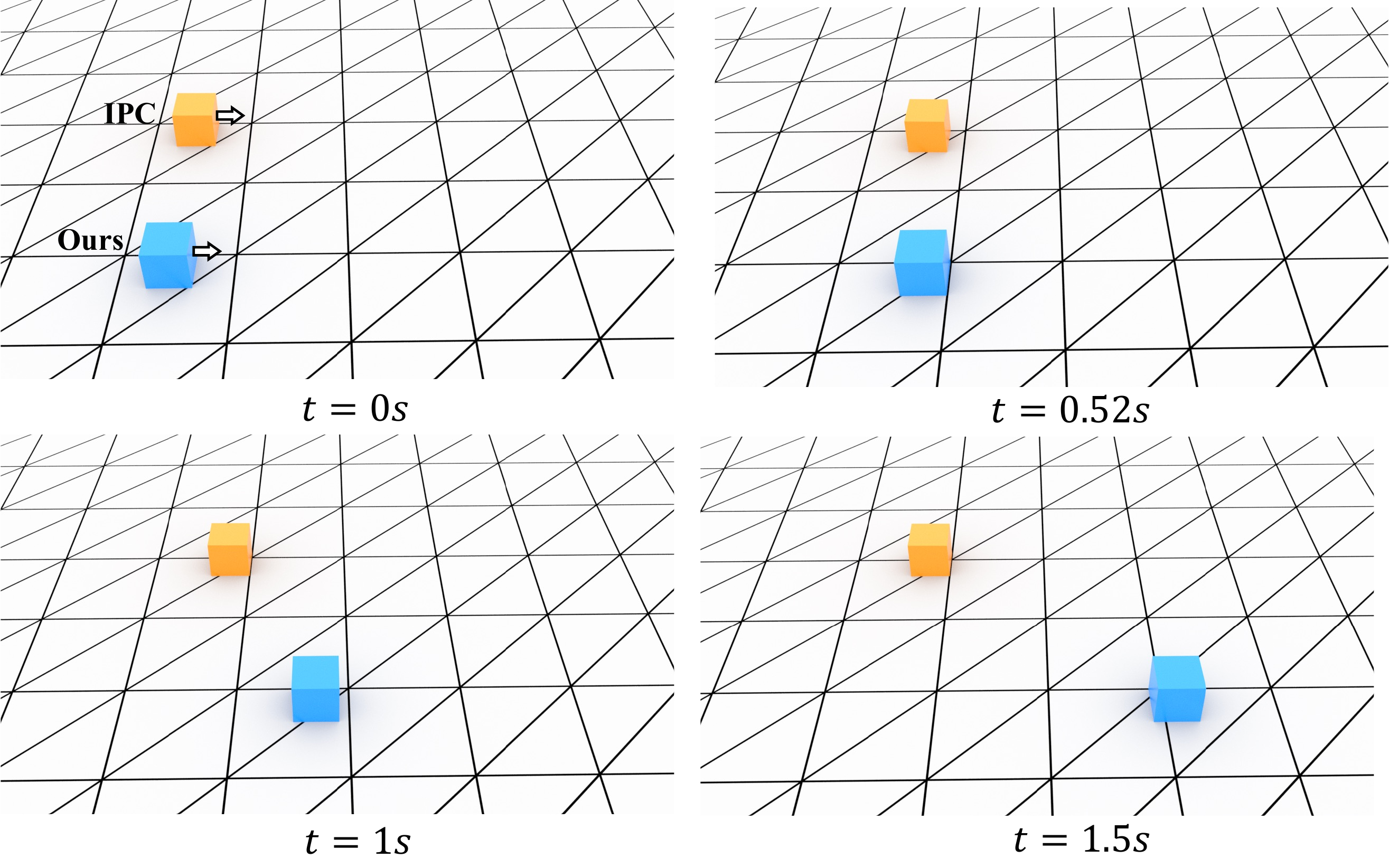}
    \caption{Comparison with IPC. A deformable cube is pressed down onto a frictionless floor and subjected to a horizontal force (shown with arrows). Whereas the energy wall from IPC induces spurious tangential forces that put an early stop to the cube's trajectory (orange), our approach produces the expected linear motion (blue).
    }
    \label{fig:ipc_failure}
\end{figure}

\paragraph{Qualitative Comparison.}
We begin our analysis by examining a forward simulation task where the spurious tangential forces generated by IPC become evident. To this end, we consider a deformable cube that is pressed down onto a frictionless surface and subjected to a constant horizontal force. The desired outcome is for the cube to move in the direction of the applied force. However, as indicated in Fig.~\ref{fig:ipc_failure} and shown in the accompanying video, IPC produces undesirable tangential forces making the cube grind to a halt. This behavior is explained by the presence of energy walls arising from the superposition of contact potentials (see also Fig. \ref{fig:IPCEnergyWall}. These energy walls generate artificial friction-like forces that prevent the cube from moving along the desired trajectory. In contrast, our approach accurately reproduces the expected linear motion.

To generate simulations with IPC, we use the publicly available code base\footnote{https://github.com/ipc-sim/IPC} from Li \etal \shortcite{Li2020}, with modifications made solely to the scene configuration. The scene description file can be found in the supplemental material accompanying this paper.
As shown in the accompanying video, these artifacts are observed consistently for different hyper-parameters, including the threshold $\hat{d}$ that controls when contact potentials are activated.

\begin{figure}[h!]
    \centering
    \includegraphics[width = \linewidth]{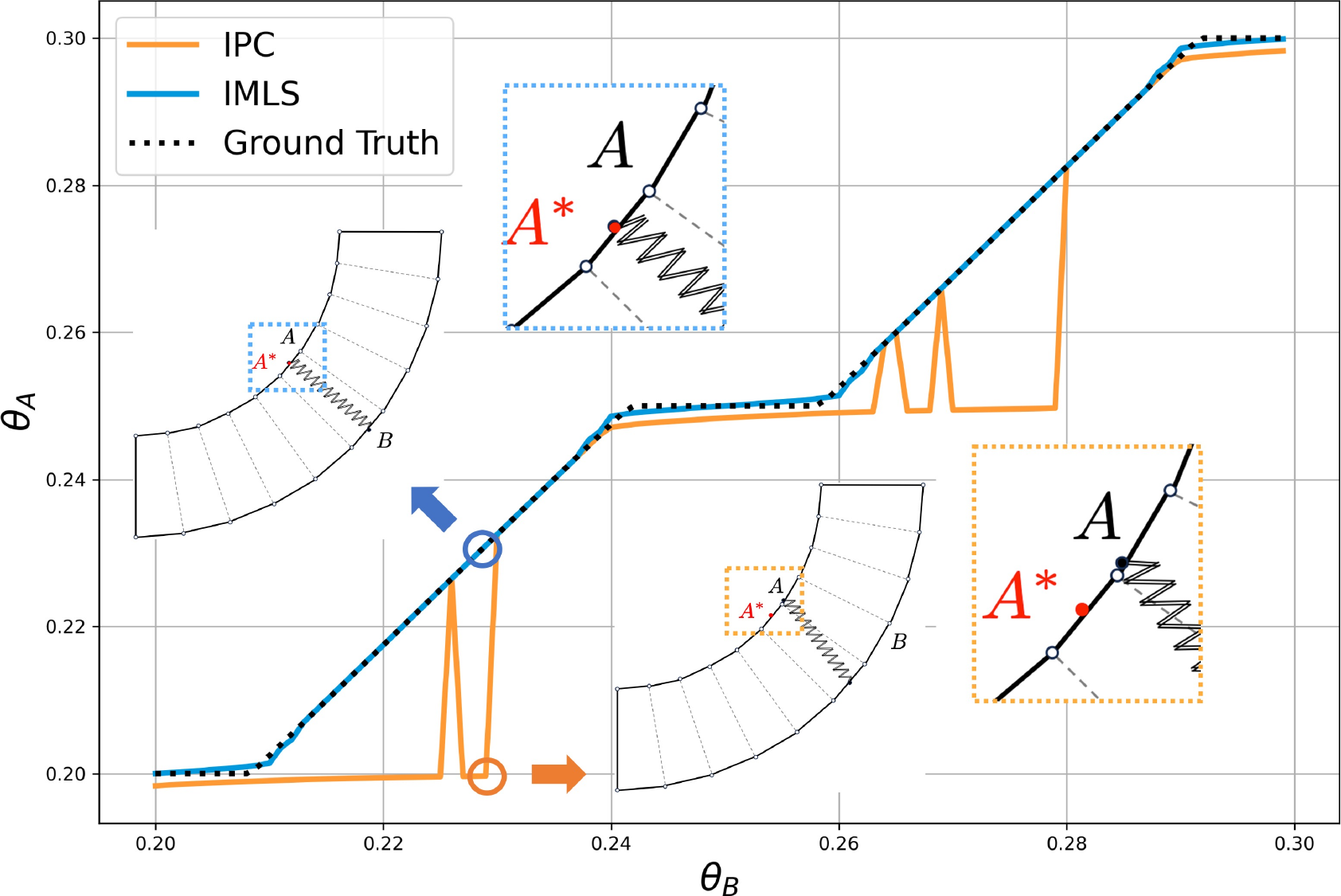}
    \caption{Accuracy of the forward problem. Our approach (solid blue line) consistently arrives at locations that closely match the ground truth values (dotted line). IPC (solid orange line) suffers from the presence of energy barriers at segment transitions. As illustrated in the inset for IPC, point $A$ fails to overcome the energy wall originating from its closest vertex, preventing it from reaching its target ($A^*$). }
    \label{fig:sliding_forward}
\end{figure}

\paragraph{Quantitative Comparison.}
We now shift our focus to an inverse design example where spurious tangential forces prevent convergence to the desired solution.
\begin{wrapfigure}{r}{0.2\textwidth} 
    \centering
    \includegraphics[width=0.9\linewidth]{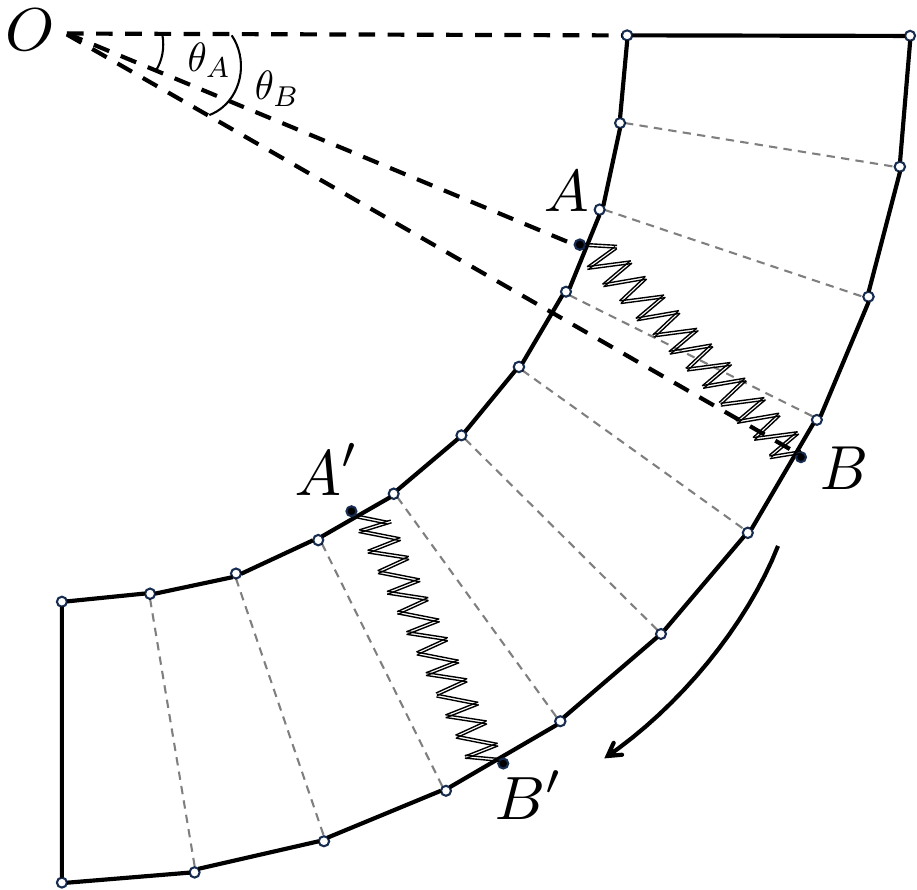}
    \label{fig:SlidingContactTestSetting}
\end{wrapfigure}
As shown in the inset figure, we examine a case where two vertices slide along a rigid body in the form of a quarter annulus. The two vertices, denoted as $A$ and $B$, are placed on the inner and outer tracks of the annulus and connected by a zero-length spring. In the absence of friction forces, moving vertex $B$ should lead to an analogous motion for vertex $A$ such as to minimize their distance. 
We first examine the accuracy of both approaches on the forward problem, where we seek the equilibrium states of $A$ when $B$ is moved to a prescribed location. 
The comparison between IPC and our approach with the ground truth solution is shown in Fig.~\ref{fig:sliding_forward}. As can be seen from this figure, our approach achieves accurate matching of the targets throughout the entire range of motion. In contrast, IPC exhibits poor matching behavior. By examining the insets in Fig.~\ref{fig:sliding_forward} (targets shown in red), we observe that this failure is once again due to spurious tangential forces stemming from energy walls. In particular, point $A$ comes to a halt at its closest vertex and fails to reach its target location.
\par
\begin{figure}[h!]
    \centering
    \includegraphics[width = \linewidth]{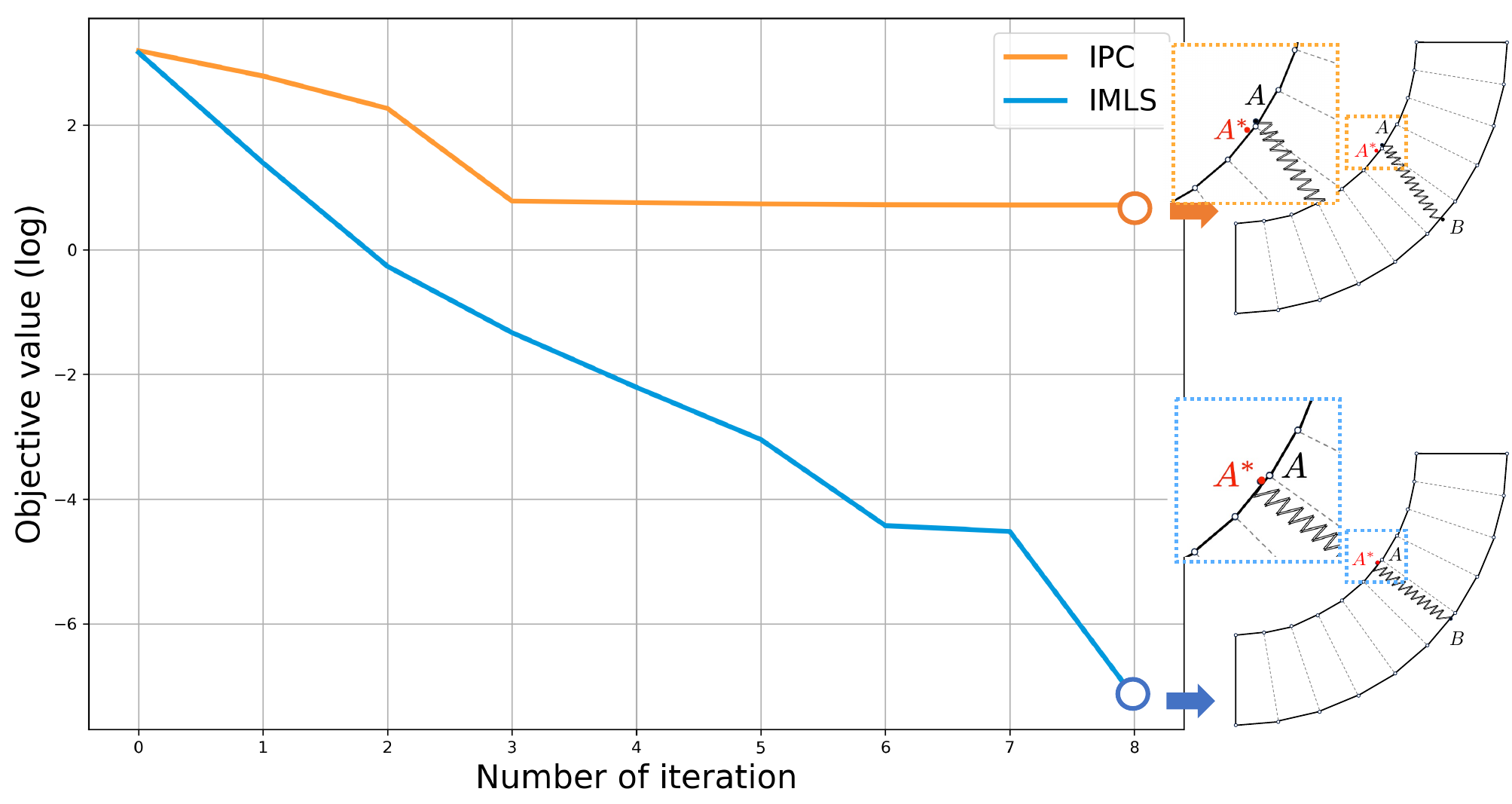}
    \caption{Accuracy of the inverse problem. The convergence plots for IPC and IMLS are shown in orange and blue, respectively. IPC fails to find the true solution to this inverse design problem due to the presence of energy walls that introduce local optima. In contrast, our approach demonstrates robust convergence, reaching the solution in just 8 steps.}
    \label{fig:sliding_inverse}
\end{figure}

Next, we consider an inverse version of the problem, where the goal is to control the position of the vertex $B$ such that vertex $A$ assumes a given target location $A^*$.
We formulate this task as a constrained optimization problem where the constraint enforces static equilibrium,
\begin{align}
\label{eqn:inverse}
    \begin{split}
        \underset{\theta_B}{\min} \quad O(\bx_B(\theta_B)) &= ||\bx_A(\theta_A)(\bx_B(\theta_B)) - \bx_A^*(\theta_A)||_2^2,\\
        &\text{s.t.} \quad \frac{\partial E}{\partial \bx} = \bZero \ ,
    \end{split}
\end{align}
where $E$ is the total energy of the system, $\bx_A^*(\theta_A)$ is the target location prescribed by a clock-wise rotation $\theta_A$ \wrt the horizontal axis in the image plane, and the positions of both points are concatenated into $\bx$. The optimization consists of a single degree of freedom $\theta_B$, and we minimize this objective (Eqn.~\ref{eqn:inverse}) using gradient descent. The simulation derivative $\frac{\dd \bx_A}{\dd \bx_B}$ is obtained using sensitivity analysis.
The convergence plot is shown in Fig.~\ref{fig:sliding_inverse}. While IPC fails to converge to the desired accuracy, our approach demonstrates robust convergence. This discrepancy is again explained by the energy walls introduced at segment transitions when using IPC, which introduce undesirable local optima. 
We note that increasing mesh resolution for IPC would change the frequency of energy walls without removing them.

\begin{figure*}[ht]
    \centering
    \includegraphics[width=\linewidth]{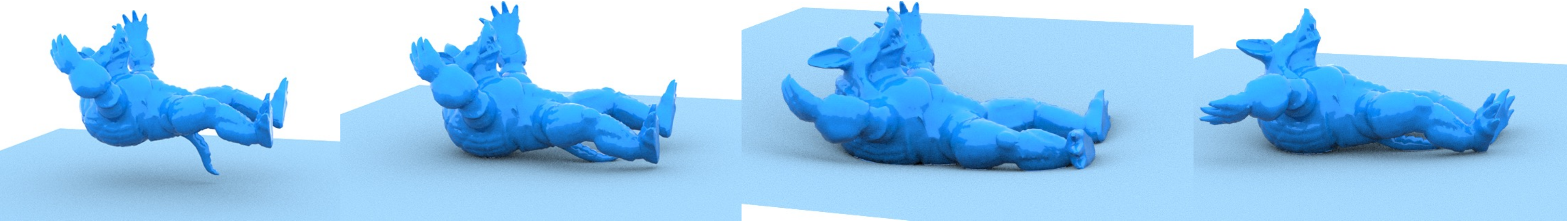}
    \caption{Free falling. A soft armadillo is falling onto a deformable plane. Both the plane and the armadillo's surface are represented using IMLS. 
    }
    \label{fig:free_fall}
\end{figure*}

\begin{figure*}
    \centering
    \includegraphics[width=\linewidth]{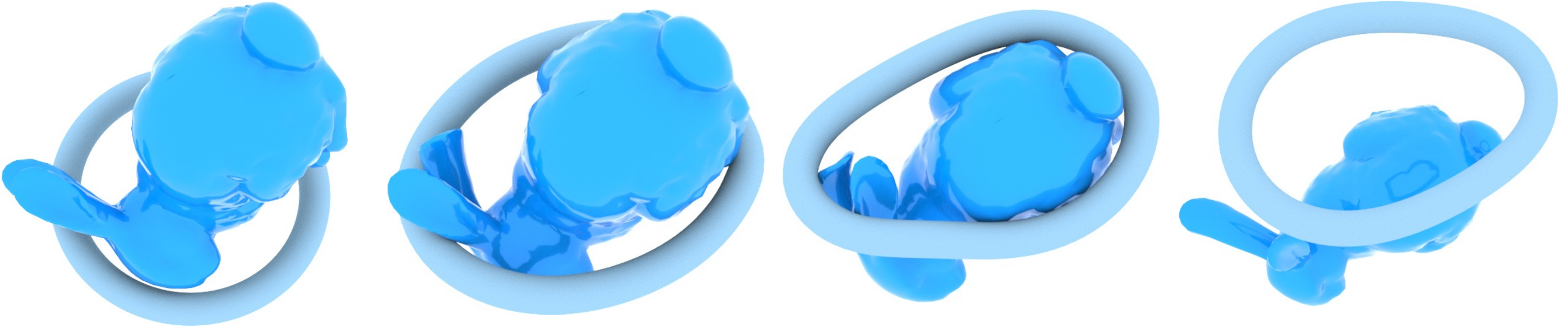}
    \caption{Torus tunneling. A soft bunny falls under gravity through a torus, with both geometries represented by implicit moving least square surfaces for contact resolution. 
    }
    \label{fig:bunny_torus}
\end{figure*}

\begin{figure*}
    \centering
    \includegraphics[width=\linewidth]{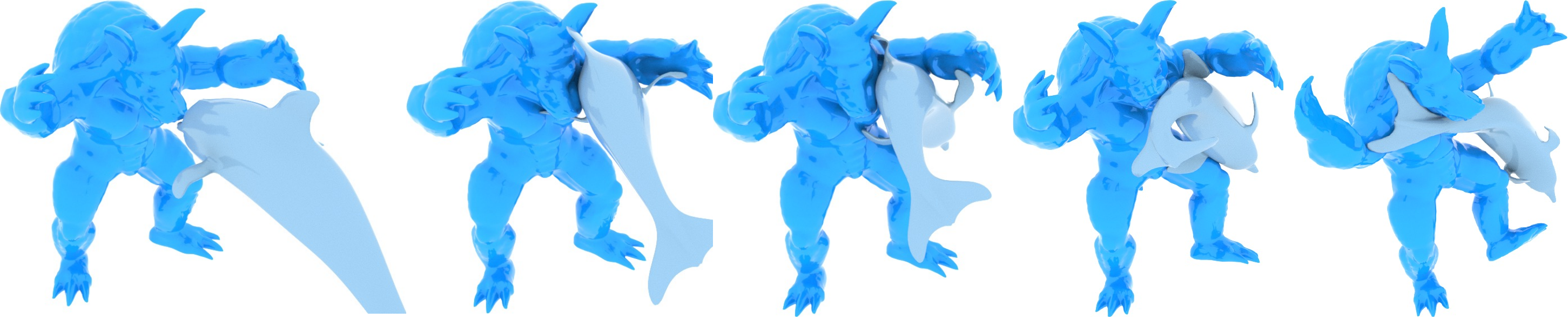}
    \caption{Catch-of-the-day sequence. Our approach robustly handles intricate contacts between deformable bodies with complex geometries.
    }
    \label{fig:catch_full}
\end{figure*}
\subsection{3D Examples}

\label{sec:result3d}
Lastly, we showcase the effectiveness of our approach through a series of 3D examples with deformable bodies represented using IMLS surfaces. In the first example (Fig.~\ref{fig:free_fall}), an armadillo lands on a soft plane. The second example (Fig.~\ref{fig:bunny_torus}) shows a soft bunny passing through a narrow deformable torus. Finally, the interactions between complex geometries can be seen in Fig. \ref{fig:catch_full}, in which a deformable dolphin collides with a soft armadillo. As can be seen from these sequences, our approach reliably handles complex contact scenarios between deformable bodies undergoing large deformations. We refer to the accompanying video for the complete simulation sequences.
\subsection{Implementation Details.}
Our framework is implemented in C++ with Eigen \cite{eigenweb} for linear algebra operations. The Eigen wrapper for the CHOLMOD solver \cite{chen2008algorithm} is used for solving linear systems. The total energy is minimized using Newton's method augmented with a back-tracking line search and adaptive regularization. We use linear tetrahedron elements for deformable solids and discrete shell elements \cite{grinspun2003discrete} for surface meshes. A standard Neo-Hookean material~\cite{bonet1997nonlinear} is used for all of our examples, and we use a step size of $0.005s$ for time integration. The timings are obtained on a workstation with a \textit{AMD Ryzen 7 3700X} CPU and summarized in Table \ref{tab:stats}.

We follow the original robust IMLS paper \cite{oztireli2009feature} using the kernel $\phi_i(x) = (1-\frac{|x-x_i|^2}{R^2})^4$, where $R$ is the support radius that ranges from $1.4$ to $4$ times the mesh resolution. $w =e^{-\frac{|f_x-f|^2}{\sigma^2_r}}$ and $w_n = e^{-\frac{|n_p-\nabla f|^2}{\sigma_n^2}}$ where $\sigma_r = 0.5$ and $\sigma_n = 1$. We use 1 iteration of optimization as \"Oztireli et al. suggested. The Young’s Modulus of the material we used in simulation is $10^6 Pa$ (similar to a soft natural rubber) and the Poisson ratio is set to 0.3. 

\begin{table}[h!]
\caption{Statistics for 3D examples. We report the number of simulation degrees of freedom, the average computational time per Newton iteration, the maximum number of vertices under contact, and the average number of Newton iterations per time step. }  
\label{tab:stats}
\begin{tabular}{ |c|c|c|c|c|c| } 
\hline
Example & \# DoFs.  & Time/Iter. & Max \# contact vtx & Avg \# Iter. \\
\hline
Fig. \ref{fig:free_fall} & 51,978  & 13.318s & 3,798 & 13.6 \\ 
\hline
Fig. \ref{fig:bunny_torus} & 32,062  & 6.349s & 1,136 & 13.0 \\ 
\hline
Fig. \ref{fig:catch_full} & 108,513  & 16.540s & 1,322 & 19.9 \\ 
\hline
\end{tabular}
\end{table}
\section{Conclusions}

We have investigated a fundamental limitation of existing methods for handling contact between deformable bodies based on $C^0$-discretizations. Our analysis showed that spurious tangential forces generated by IPC-type approaches can significantly affect simulation outcomes and prevent inverse problems from converging to the right solution. 
We furthermore showed that using smooth surface representations based on IMLS effectively resolves this problem, leading to robust behavior even for challenging contact scenarios and complex geometries.

\subsection{Future Work}
%
While we have demonstrated that our smooth representation enables robust convergence for inverse design on a 2D example, applications of our approach to inverse design tasks in 3D deserve further investigation.
We have focused on frictionless contact to isolate the impact of spurious tangential forces. While these effects might be less pronounced for high-friction scenarios, we argue that accurate modeling of normal contact forces is a prerequisite for accurate frictional behavior. Nevertheless, exploring the impact of spurious tangential forces in the context of frictional contact is an interesting direction for future work.
%


\begin{acks}
This work was supported by the European Research Council (ERC) under the European Union’s Horizon 2020 research and innovation program (grant agreement No. 866480), and the Swiss National Science Foundation through SNF project grant 200021\_200644.
\end{acks}
\bibliographystyle{ACM-Reference-Format}
\bibliography{reference}
\end{document}